\documentclass[conference]{IEEEtran}
\usepackage[pdftex]{graphicx}
\DeclareGraphicsExtensions{.pdf,.jpeg,.png,.gif,}
\usepackage{amsmath}
\usepackage{url}
\ifCLASSINFOpdf
\else
\fi
\usepackage{color}
\usepackage{graphicx}
\usepackage{tabularx}
\usepackage[export]{adjustbox}
\usepackage{siunitx}
\usepackage[english]{babel}
\usepackage{cite}
\usepackage{enumitem}
\usepackage{multirow}
\usepackage{pbox}
\usepackage{amsmath}
\usepackage[euler]{textgreek}
\usepackage{booktabs}
\newcolumntype{P}[1]{>{\centering\arraybackslash}p{#1}}

\usepackage[nolist]{acronym}
\usepackage{glossaries}
\usepackage{algorithm}
\usepackage{algorithmic}
\usepackage{epsfig}
\usepackage{wrapfig}
\usepackage{xcolor}
\usepackage{setspace}
\usepackage{multirow}

\usepackage{flushend}
\usepackage{times,amssymb,amsmath,amsthm,epsfig,subfig,graphicx,algorithm,algorithmic,url}
\usepackage{multibib}
\usepackage{multirow,bbm}
\usepackage{booktabs}
\usepackage{verbatim}
\usepackage{float} 

\def\rrr#1\\{\par
\medskip\hbox{\vbox{\parindent=2em\hsize=6.12in
\hangindent=4em\hangafter=1#1}}}

\begin{document}
\bstctlcite{IEEEexample:BSTcontrol}
%
\title{Security Issues of Low Power Wide Area Networks in the Context of LoRa Networks}
%
%
%

\author{\IEEEauthorblockN{Debraj Basu, Tianbo Gu and Prasant Mohapatra}
\IEEEauthorblockA{Department of Computer Science, University of California, Davis, CA, USA\\
 Email: \{dbasu, tbgu, pmohapatra\}@ucdavis.edu,
}

}

\maketitle
\begin{abstract}

Low Power Wide Area Networks (LPWAN) have been used to support low cost and mobile bi-directional communications for the Internet of Things (IoT), smart city and a wide range of industrial applications. A primary security concern of LPWAN technology is the attacks that block legitimate communication between nodes resulting in scenarios like loss of packets, delayed packet arrival, and skewed packet reaching the reporting gateway. LoRa (Long Range) is a promising wireless radio access technology that supports long-range communication at low data rates and low power consumption. LoRa is considered as one of the ideal candidates for building LPWANs. We use LoRa as a reference technology to review the IoT security threats on the air and the applicability of different countermeasures that have been adopted so far. LoRa extends the transmission range by controlling the spreading factor (SF) and in turn, the data-rate. LoRa nodes that are close to the gateway use a small SF than the nodes which are far away. But it also implies long in-the-air transmission time, which makes the transmitted packets vulnerable to different kinds of malicious attacks, especially in the physical and the link layer. Therefore, it is not possible to enforce a fixed set of rules for all LoRa nodes since they have different levels of vulnerabilities. Our survey reveals that there is an urgent need for secure and uninterrupted communication between an end-device and the gateway, especially when the threat models are unknown in advance. We explore the traditional countermeasures and find that most of them are ineffective now, such as frequency hopping and spread spectrum methods. In order to adapt to new threats, the emerging countermeasures using game-theoretic approaches and reinforcement machine learning methods can effectively identify threats and dynamically choose the corresponding actions to resist threats, thereby making secured and reliable communications.

\end{abstract}


\IEEEpeerreviewmaketitle

\section{Introduction}
\label{Intro}
According to the estimation of Boston Consulting Group~\cite{IoTMarket}, \$267 Billion will be spent on Internet of things (IoT) technologies, products, and services by 2020. For example, the IoT market size of smart cities is predicted to grow to \$147.51 Billion by 2020 \cite{MarketsandMarkets}, since IoT can solve many critical issues currently faced by urban cites, like high energy consumption~\cite{shrouf2015energy, wang2016green}, environment pollution \cite{xiaojun2015iot, hromic2015real} and transportation congestion \cite{kyriazis2013sustainable, rizwan2016real}.
An efficient networking system is essential to achieve real-time monitoring and intelligent control of physical objects for IoT applications. Security and privacy of the data that are exchanged between the network entities are an integral part of an efficient networking system and are usually defined by confidentiality, integrity, and availability (CIA) in the information security paradigm \cite{CIA2017,CIA2016}. 
\subsection{IoT networks and security}
The majority of IoT enabled technologies are directed towards building smart infrastructure projects \cite{IoTProjects}. The Array of Things (AoT) \cite{AoT} is an urban sensing network of programmable, modular nodes that are deployed around cities to collect real-time data on the city's environment, infrastructure, and activity for research and public use. The Chicago Park District maintains sensors in the water at beaches along Chicago's Lake Michigan lakefront \cite{ChicagoWater}. These sensors capture the measurements at a periodic rate along the lakefront.  Pervasive Nation is Ireland's Internet of Things testbed operated by CONNECT, headquartered at Trinity College Dublin, the University of Dublin. This testbed is built on LPWAN technology enabled by LoRa \cite{PervasiveNationIreland}. There are other railroads and port infrastructure projects that are leveraged by IoT where hundreds of sensors are deployed \cite{IoTProjects}.
The IoT network communication channel is predominate wireless; thus, the on-air legitimate control and data messages can be overheard and modified by an attacker. Moreover, an attacker with malicious intent may inject illegitimate messages into the network. In \cite{MehmoodIoTChallenges2017}, authors have identified security, privacy, and trust as the significant challenges to build smart city projects. It has been pointed out in the white-papers by WIND RIVERS \cite{WINDswhitepaper2017} that IoT security is more challenging than cybersecurity because of the large attack surface presented by the millions of IoT devices. Most of these devices are resource-constrained and therefore, limited by computing power for encryption capabilities. They are also expected to operate for years without being replaced, hence prolonging their exposure to attack from newer attack vectors.

Recently, Low-Power Wide Area Networks (LPWANs) have emerged as an attractive communication technology for IoT~\cite{moyer2015low}.
They can support large-scale coverage with long communication distance at low cost and long network lifetime. In an LPWAN network, all sensor devices directly transmit data to an LPWAN gateway and can work for years with low energy consumption. An LPWAN gateway covers a large area of many miles and thousands of sensor devices. The collected data is transmitted and stored in a network server to be processed by different applications. 

At present, four LPWAN technologies are mainly available, i.e., LoRa~\cite{LoRa}, SigFox~\cite{SigFox}, Narrowband Internet of Things (NB-IoT)~\cite{NBIoT} and Long Term Evolution for Machine type communication (LTE-M)~\cite{LTEM}. NB-IoT has been developed and standardized by 3GPP. NB-IoT is designed to support very low power consumption and low-cost devices in extreme coverage conditions \cite{NBIoT2017}. NB-IoT and LTE-M work on licensed cellular frequency bands, which are implemented on existing cellular infrastructures by mobile operators. LoRa and SigFox work on the unlicensed 900 MHz band but adopt different business models.

\subsection{LoRaWAN and unique security threats}
LoRaWAN is the standard for wireless communication protocols that allows IoT devices to communicate over large distances with minimal battery usage. LoRaWAN supports single-hop network topology with the end devices connected to the network servers via intermediate gateways. The communication between the LoRa enabled sensor nodes, and the gateways go over the wireless channel utilizing the LoRa physical layer, while the connection between the gateways and the central server are handled over a backbone Internet Protocol(IP)-based network \cite{LoRaNetworkArch}. The data rate of LoRa networks is determined by the spreading factor (SF). Higher SF corresponds to lower data rates and in turn, long-range communication. A very low data rate (0.3-50 kbps depending on the frequency band) enables LoRa to cover long-range communication (1-2 miles). However, it also increases the on-air transmission time in the order of 3 seconds (depending on the payload).  This is a \textit{unique feature of LoRa network} when the security risks of the nodes in the network are not the same even when these nodes use single-hop for communication.  Therefore, it makes LoRa network susceptible to different kinds of \textit{Denial of Service (DoS)} attacks, including jamming, replay attacks, and eavesdropping and in different variability depending on the distance of the LoRa nodes from the gateway. 

DoS attacks are launched to destroy the availability of one or multiple nodes. For example, an adversary may physically tamper an end device to disable its duty or make it transmit interference signals, or it can sniff the packets in wireless channels and selectively jam a particular portion of end devices. DoS attacks disrupt legitimate transmissions and increase the energy consumption of end devices by more re-transmissions. 

In a jamming attack, a high power transmitter can transmit small packets either continuously or randomly and interfere with legitimate packet transmissions, disrupting the regular network operation \cite{tomic2017survey}. Experiments done with vehicular communication have shown that RF jamming can lead to large communication-blind areas \cite{PunalJamming2012}. Not much can be done to mitigate jammers with unlimited resources in terms of transmission power and spectrum efficiency \cite{PelechrinisJamming2009}. LoRa works in the frequency band of 26 MHz in the USA (902 to 928 MHz). Such a narrow band is not immune to such kind of wideband jamming attacks. It is, therefore, essential to seek the support of law enforcement to capture the attacker physically. However, a continuous or wide-frequency-band jamming attack is easy to detect. Usually, an attacker would not reveal its presence but only listens to the channel passively. It selectively jam packets by reading the physical header and go to sleep or in listening mode after jamming the packet \cite{shiu2011physical}. This attack mode is hard to detect and deter. In LoRa, since the on-air time of packets is high, the reaction time of the selective jamming attacker to jam the packet after the header is read is also high. Therefore, this kind of attack is a significant threat to the security of the LoRa network. On the top, LoRa end-devices use random time slots to transmit packets. Therefore LoRa network cannot distinguish between packet losses due to regular congestion and a jamming attack.

In replay attacks, a valid transmission is repeated by the malicious attacker, generating false messages to the gateway and denying legitimate messages to reach the network server or rejects a valid \textit{network-join request} \cite{avoine2017rescuing}. To maintain the integrity of valid network join requests, LoRa uses random numbers (DevNonce) that are derived from the physical layer signal strength values. However, an attacker can destroy the randomness by injecting high power packets in the network. Such an attack has been a critical security issue for nodes using LoRaWAN v.1.0. Some existing works in \cite{TomasinSLoRaVulnerability2017, vanEs2018, NaSeungReplayAttackLoRa2017, mastersthesisZulian2016} have analyzed the replay attack scenarios in the context of the LoRa network when a node attempts to join the network. 

 When the network server receives a join request message, it checks if the DevNonce is used from the pool of last DevNonce ($N_{D}$) not used. If the number matches, the join request is rejected. For a given number of join requests per day per device, a higher value of $N_{D}$ means that an attacker has to wait for a longer time to use a DevNonce. In paper \cite{tomasin2017security} the authors have experimentally shown using LoRa hardware SX1272 \cite{SX1272} that random number generator (RGB) can be comprised in the presence of a jammer which can make the LoRa end-device to repeat DevNonce. In the SX1272 Radio Frequency chip, the N-bit random number is obtained from the LSB (least significant bit) of the register RegRssiWideband (address 0x2c). It is assumed that the LSB continually changes due to noise and radio channel behavior and therefore used a source of random number generator. A high power jammer can make the register value constant, and the DevNonce value can no longer be random.

In LoRaWAN v.1.1 specification, DevNonce is a counter starting at 0 when the device is initially powered up and incremented with every \textit{Join-request}\cite{LoRaNetworkArch}. The specification further states that "a DevNonce value \textbf{shall never} be reused for a given join request value (\textit{JoinEUI}). If the end-device can be power-cycled, then DevNonce should be persistent (stored in non-volatile memory). Resetting DevNonce without changing the \textit{JoinEUI} may cause the Network Server to discard the Join-requests of the device". 
Since a loss of power or reset could happen at any time, LoRa end nodes that comply with LoRaWAN v.1.1 standard version must have non-volatile memory like electrically erasable programmable read-only memory (EEPROM). For each end-device, the Network Server keeps track of the last DevNonce value used by the end-device and ignores Join-requests if DevNonce is not incremented. In this way, a replay attack during a network join-request is prevented in LoRaWAN v.1.1. However, there is no reference to know the current extent of usage of the LoRa devices that comply with LoRaWAN v.1.1 and their network deployment scenarios. It remains a faulty key-management issue for LoRaWAN v.1.0 end nodes until the hardware is not replaced completely ~\cite{MundtLoRaWAN2018}.

An eavesdropper can also launch a replay attack. An eavesdropper can overhear a wireless transmission and get access to sensitive or private information. Eavesdroppers use passive receivers that only listen to the channel and hardly transmit any signal making them very difficult to detect. In order to avoid a replay attack, a nonce (random number) is used in a message to verify its freshness \cite{SyversonReplay1994}. LoRa uses a frame counter as a nonce to generate the encrypted message using the shared key between the transmitter and the receiver. When the frame counter resets but the key remains the same, an eavesdropper can capture consecutive packets to derive the plaintext \cite{avoine2017rescuing}. An eavesdropper can further launch a replay attack or selectively jam valid messages. 

Attack models are hard to predict, and attackers are equally difficult to detect. It should be a two-prong countermeasure to detect any potential threat and progressively learn about the threat model (e.g., how it is affecting the confidentiality, integrity, and availability of the network). In this way, any countermeasure becomes more confident and robust. Different consistency checks of network parameters can be used to detect an attack. The game-theoretic approach can be adopted with two adversaries trying to maximize their utility functions \cite{ManshaeiGameThoery2013}. Reinforcement learning methods are useful when the attack model is not known in advance. 
\subsection{Survey papers and our contributions}
Recent survey papers have done security risk analysis of LoRaWAN \cite{donmez2018security,butun2019security,YangLoRaWAN2018, vanEsLORAWAN2018, SaariMLORA2018, aras2017exploring}. However, none of these papers have addressed the security threats that are unique to LoRa networks. There is no clear strategy to identify and detect a threat, and countermeasures when the threat models are unknown. 
Overall, we did not see security solutions that cater specifically to the LoRa network.

Our paper has done an exhaustive survey of existing solutions that can be applied to LoRa networks. Our paper has identified external jamming and faulty key management as two primary sources of security threats to LoRaWAN. We have covered all major security approaches to counter jamming, replay, and eavesdropping in the context of the LoRa network. 

Our analysis shows that game-theoretic and reinforcement learning approaches can be used to counter these security threats, primarily when the attack characteristics are not known. These proposed approaches can take advantage of the state-of-the-art classification algorithms to classify threats and tune the transmission parameters that suits best to counter the attacker. In the context of LoRa, it can use tunable SF to control the data rate and therefore, on-air time to reduce its vulnerability. 
Hence there is a vast scope of further research, and our paper has proposed possible future research directions to make the LoRa network more secure.

We have organized the remaining of this paper in the following order. 
Section \ref{LoRaTech} has discussed the general LoRaWAN technological and security features. We discuss the vulnerability issues in a wireless network in general in section \ref{wirelessNetwork} to get a broad understanding of threats. Sections \ref{UniqueLoRa} discuss the unique features of LoRa and the threats in detail. Finally, we explore the solutions that apply to LoRa networks in Section \ref{LoRaSolutions}.  

\section{LoRaWAN as Low Power Wide Area Network (LPWAN) standard}\label{LoRaTech}

According to rfc8376 \cite{rfc8376}, 'Low-Power Wide Area Networks are wireless technologies with characteristics such as large coverage areas, low bandwidth, possibly very small packet and application-layer data sizes, and long battery life operation.'


\subsection{LPWAN and security}
Paper\cite{LPWASurvey2017, LPWAsurvey2017whitepaper, raza2017low} have surveyed the different LPWAN technologies and their security features. The design goals for successful deployment and operation of LPWAN technology are long-range, ultra-low power operation, low cost, scalability, and quality of service. The extended range for wide area coverage is achieved by operating in sub GHz range for low signal attenuation as compared to GHz bands. It also uses spread spectrum (SS) techniques to be resilient to interference and robust to jamming attacks. To achieve a long battery life, LPWAN technologies use single-hop communication to avoid high deployment cost of mesh networks and congestion due to the network traffic pattern. They also control the sleep/wake-up schedules (duty cycle) to minimize energy usage. 
 
 When it comes to accessing the common media (usually wireless between the end nodes and the Gateway), carrier sense multiple access with collision avoidance (CSMA/CA) \cite{chen2007analytical} is the most popular media access control protocol (MAC). This channel access protocol has been successfully deployed in WLANs and other short-range wireless networks. The virtual carrier sensing using Request to Send/Clear to Send (RTS/CTS) mechanism inflicts excessive signaling overhead and is usually avoided in LPWAN technologies. However, the use of random access methods can make LPWAN vulnerable to deliberate jamming attacks that increase the packet collision rate and decrease the overall network throughput.
  \begin{table*}[h]
  \centering
  \caption{Device classification in LoRaWAN \cite{LoRaNetworkArch, LoRaClass, LoRaClassServer}}
  \label{tab:LoRaWANclass}
  \begin{tabular}{|p{15em}| p{15em}| p{15em}|}
 \hline
 \textbf{LoRa Class A} & \textbf{LoRa Class B} & \textbf{LoRa Class C}\\
    \hline
   Asynchronous and bidirectional & Synchronous and bidirectional & Synchronous and bidirectional\\
  \hline
   In sleep mode except when transmitting & Open downlink receive slots at scheduled times & In receive mode except when transmitting \\
   \hline
  Initiates uplink communications & Synchronized with periodic beacons & Network server can initiate uplink transmission at any time\\
   \hline
   No guaranteed latency & Fixed downlink latency & No latency \\
   \hline
   Lowest power operating mode &  Additional energy consumption due to beacon signal transmissions &  Significant power drainage in the order of 50 mW \\
  \hline
        \end{tabular}
\end{table*}
In terms of network security, due to cost and energy limitations, LPWAN usually avoids cellular type authentication, security, and privacy mechanisms. Most LPWAN technologies use symmetric key cryptography to authenticate end devices with the network and preserve the privacy of application data. However, over-the-air (OTA) security features, including authentication, are not well supported in LPWAN technologies and can expose the end-nodes to threats over a prolonged duration, including eavesdropping and replay attacks. 

 \begin{figure}[h]
    \centering
    \includegraphics[scale=0.31]{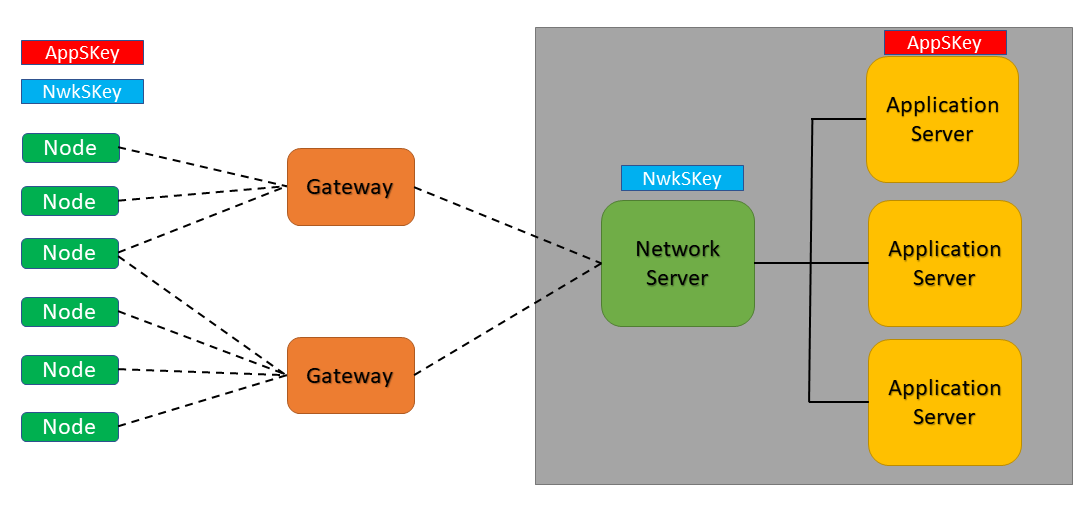}
    \vspace{0cm}
    \caption{\textbf{High level LoRa network architecture \cite{LoRaNetwork2016}}}
    \label{fig:LoRaArchitecture}
\end{figure} 
Like a cellular service provider, SigFox deploys its gateways in some cities and users subscribe to its service \cite{Sigfox2017}. On the contrary, LoRa is open-standard. It has released open-source hardware (except the chips) and software. Users can build their autonomous LPWAN networks. However, the available open-source hardware and software of LoRa networks only provide some essential functions, e.g., single-link transmissions and ALOHA-based multiple access. 

Sigfox application payload is not encrypted, while in LoRaWAN, a unique 128-bit encryption key is shared between the end-device and network server, and another unique 128-bit key is shared end-to-end at the application level.


\subsection{LoRaWAN and security}
 

LoRaWAN supports three different classes (Class A, B, and C) of LoRa devices to address the diverse application needs. They are tabulated in Table \ref{tab:LoRaWANclass}.

Class A operation is the most popular and preferred mode because of the lowest energy consumption among all the three classes. End devices choose random slots to transmit packets. Most of the devices are battery powered and maximize energy consumption by going to sleep mode after transmission. 

 A high-level schematic diagram of the LoRa network is shown in Fig. \ref{fig:LoRaArchitecture}. Each end device can connect to multiple gateways that are, in turn, linked to the network server. The network server can connect to multiple application servers.

As per LoRaWAN security documentation by LoRa Alliance \cite{LoRaWANSecurity_whitepaper2017} and The Things Network \cite{ThingNetLoRa2018,iskhakov2017analysis}, LoRaWAN uses the symmetric-key Advanced Encryption Standard (AES) \cite{HeronAES2010} for encrypting, decrypting and message signing. To ensure confidentiality of a message from the end devices to the application servers, LoRaWAN (since v1.0.2) encrypts the payload in a message with the 128-bit application session key (AppSKey). For integrity, all LoRaWAN messages are signed with a Message Integrity Check (MIC). It is derived from a hash of the message content and a 128-bit network session key (NwkSKey). The AppSkey and NwkSKey are derived from a root key (AppKey). The end-device and the network server know this key. All keys are unique for each end-device. For availability, LoRaWAN supports multiple gateways (combined with packet de-duplication on the network server) \cite{vanEs2018}. 

\begin{figure}[t]
    \centering
    \includegraphics[width=\linewidth]{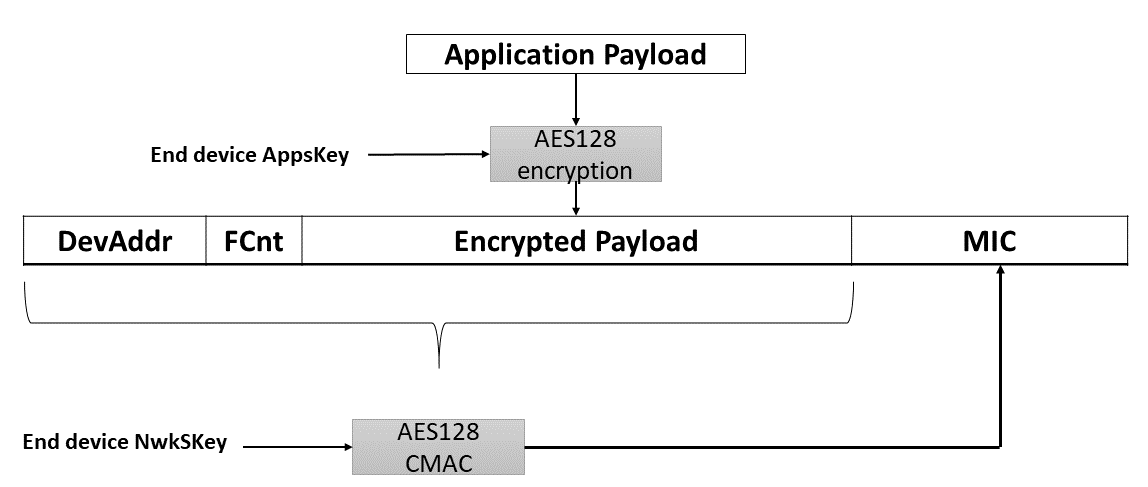}
    \caption{\textbf{Encryption of payload and message signature \cite{mastersthesisZulian2016}}}
    \label{fig:EncryptionDiagram}
\end{figure}

Fig. \ref{fig:LoRaArchitecture} illustrates the usage of the two keys to maintaining the confidentiality and integrity of the LoRa networks. The basic idea is that communication should be secure on multiple levels. A network server does not need to be able to read the actual contents of the message if it is not relevant for the network or infrastructure. Therefore, NwkSkey and AppSKey are used during normal message exchange. Fig. \ref{fig:EncryptionDiagram} shows a simplified diagram of the use of AppSKey and NwkSKey. The AppSKey must be only known by the end-device and application server, whereas, the NwkSKey, must be known by the end-device and network server only. The frame counter, $FCnt$ is maintained and evaluated for MIC in both the end device and the network server to avoid replay attacks.

LoRa end devices are also vulnerable to replay attacks during activation or network join sessions. Each end device must join the LoRaWAN network before it can become active in the network. Each end device has a unique 128-bit app key (AppKey), and that is used when the node sends a join-request message. The join request message is not encrypted but signed using the AppKey. 
There are two ways to join a LoRaWAN, i.e., Over-the-air Activation (OTAA) and Activation-by-personalization (ABP).

In the OTAA method, the join-request message contains the AppEUI and DevEUI which are unique to the owner of the device and globally, respectively. The message also contains a DevNonce which should be a randomly generated two-byte value, and the network server keeps track of this number to avoid a replay attack. A 4 Byte MIC using the AES128 CMAC process is used to maintain the integrity of these three values. At the network server end, the MIC is recalculated and checked using the AppKey that is already known to the server. After validation, the network server generates its nonce value (AppNonce) and calculate the end-nodes app session key (AppSKey), and the network session key (NwkSKey) that will be used for further communication between the node and the network servers. The node receives the join-accept message after encrypted with the AppsKey from the server. The node uses the AppKey to decrypt the message and generate the AppSKey and NwkSKey key using the AppNonce \cite{MillerLoRa2016}.

In ABP, the end nodes come with the DevAddr. Both the session keys (NwkSKey and AppSKey) should be unique to the node. The end node can directly start communication with the network server without any need of join messages.

\section{Wireless networks and security} \label{wirelessNetwork}
\label{Broadthreats}
LPWAN is a network of IoT sensors that communicate over wireless channels to the reporting gateway. These gateways connect the nodes to network servers and eventually to the different application servers \cite{LoRaNetworkArch}. Therefore, the common security threats in the different layers of the wireless sensor network apply to LPWAN \cite{LPWANThreats2018}. The wireless nature of communication and resource restrictions make these sensor networks susceptible to many attacks. A comprehensive list of attack types and their implications on network performance can be found in \cite{tomic2017survey}. They are listed in Table \ref{tab:ATtackTable}.
\begin{table*}[ht]
  \centering
  \caption{Attack types in different layers and implication on network performance}
  \label{tab:ATtackTable}
  \begin{tabular}{|l| l|l |l|}
  \hline
    \textbf{Attack type} & \textbf{Layer} & \textbf{Attack features} & \textbf{Implication on network performance}\\
    \hline
  	Eavesdropping & Physical  & Overhear and intercept data &  \multicolumn{1}{m{5cm}|}{Gain access to sensitive/private information}\\
  	\hline
  	Jamming & Physical & \multicolumn{1}{m{6cm}|}{Intentional radio transmission to disrupt communication} & \multicolumn{1}{m{5cm}|}{Cut-off communication, causing congestion, exhausting energy}\\
  	\hline
  	Collision & Data link & \multicolumn{1}{m{6cm}|}{Using busy channel to cause collision} & \multicolumn{1}{m{6cm}|}{Disrupt communication, increase interference and collision}\\
  	\hline
  	Replay attack &  Network & Repeat a valid data transmission & \multicolumn{1}{m{6cm}|}{Generate false messages, increase congestion} \\
  	\hline
  	Wormhole & Network & \multicolumn{1}{m{6cm}|}{Create low latency tunnel between two malicious nodes} & \multicolumn{1}{m{6cm}|}{Sending false or out-dated data}  \\
  	\hline
  	Node tampering &  Physical & Physical access of the end-device &  \multicolumn{1}{m{6cm}|}{Alter sensitive information (e.g. cryptographic keys, routing table)}\\
  	\hline
  	Selective jamming & Data Link & \multicolumn{1}{m{6cm}|}{data packets are selectively targeted based on policy or rules} & \multicolumn{1}{m{6cm}|}{Cut-off communication, causing congestion, exhausting energy}\\
  	\hline
  	\multicolumn{1}{|m{2cm}|}{Acknowledgement spoofing} & Data link & Create false information & \multicolumn{1}{m{6cm}|}{Delay transmission, knock out new uplink transmission}\\
  	\hline
  	Man-in-the-middle & Multi-layer & \multicolumn{1}{m{6cm}|}{Sniff network to intercept communication between nodes, for example, during key exchange stage} & \multicolumn{1}{m{6cm}|}{Gain access to sensitive or private information}  \\
  	\hline
  	Denial of service & Multi-layer & \multicolumn{1}{m{6cm}|}{A general attack type that can include multiple attacks happening simultaneously} & \multicolumn{1}{m{6cm}|}{Disrupt normal operation of the network}\\
  	\hline
  	  \end{tabular}
\end{table*}
\subsection{Eavesdropping}
Eavesdropping can be considered as an attack against WSNs when an adversary node overhears the transmissions among the sensor nodes. An eavesdropping attack is a breach of confidentiality. Such an attack is countered by encrypting messages with cryptographic keys.
\subsection{Jamming attack}
As discussed earlier, a jamming attack causes severe disruption of on-going communication between end-devices and the gateway. It is an example of a DoS attack. A jammer continuously emits radio signals, without following any medium access control rules \cite{xu2006jamming}.
\subsection{Collision}
 In a collision attack, an attacker node does not follow the medium access control rules and cause collisions with the neighbor node's transmissions by sending a short packet \cite{ReindlCollision2010}. This attack can cause a lot of disruptions to the network operation, including increasing the collision rate and end-to-end delay. Due to the inherent broadcast nature of wireless networks, it is not trivial to identify the attacker.
\subsection{Replay attack}
In the replay attack, valid data transmission is repeated or delayed with malicious intent. The message is correctly encrypted, and the receiver may treat as a correct request and take action as intended by the intruder. One method to avoid a replay attack is that both sides agree to create a random session key for a specific period before starting any communication. Instead of session keys, time-stamps in all messages are also used. In this case, the receiver accepts messages that have not been sent too long ago. The other technique is to use one-time passwords for each request \cite{ReplayAttackDef}. The attacker might either have eavesdropped a message between two sides or may know the message format from his previous communication. This message may contain the secret key for authentication.
\subsection{Wormhole attack}
In the wormhole attack, an attacker records packets (or bits) in one location in the network, tunnels them to another location and re-transmits them there into the network \cite{ChunWormhole2006}. A wormhole attack is possible even if a node has fulfilled all the authenticity and confidentiality of the communication. Wireless ad-hoc networks are most susceptible to wormhole attack as there is no central controller that manages the communication between the nodes. In this way, it can fake a route that is shorter than the original one within the network. Wormhole attacks can confuse the routing mechanism that relies on the knowledge about distances between nodes. Although the wormhole attack is typical of multi-hop networks, in combination with a replay attack, two malicious devices can act as a sniffer and a jammer. The sniffer notifies the jammer through a low latency network, and the jammer stops the packet from reaching the intended gateway.   
\subsection{Node tampering}
In this kind of attack, an adversary can gain full physical control over some wireless sensor nodes by direct access. A node-tampering is a typical attack scenario when sensor nodes are geographically spread and are usually unattended. This type of attack is fundamentally different from gaining control over a sensor node remotely by breaching or take advantage of loopholes in the security shortcomings \cite{becher2006tampering}. By gaining access to these nodes, the attacker can analyze and change the functioning of the node, learn secret key material (e.g. cryptography), alter sensor readings, control the radio function of the node (read, modify, delete, and create radio messages) without accessing the program or the memory of the sensor node. Although 'tamper-resistant' nodes can secure communication, it will incur tremendous cost on network deployment \cite{Jokhiotamper2012}.
\subsection{Selective Jamming}
In selective jamming, the scenario can be that two nodes communicate over a wireless channel when a jamming node eavesdrops the message and classify it by receiving only its first few bytes. Now this jammer can corrupt the message by interfering with its reception at the receiver. Selective jamming is also an example of a DoS when legitimate messages are stopped from reaching the intended receiver \cite{proano2010selective}.
\subsection{Spoofing attack including acknowledgment spoofing}
A spoofing attack is an example of DoS attacks where the attacker can forge its identity to launch, inject false data packets, advertise and disclose false services to other wireless nodes including false routing and control information to disrupt the wireless network operation \cite{PintoSpoofing2018}. In acknowledgment (ACK) spoofing, an attacker can capture an ACK packet, delay its transmission while selectively acknowledge an unrelated message which may not have arrived at the network server. 
\subsection{Man-in-the-middle (MITM)}
A Man-in-the-middle attack happens when an outside attacker intercepts a communication between two network entities. Eavesdropping is an example of a MITM attack.  An MIIM attack allows the attacker to intercept, send and receive messages meant for someone else without the knowledge of the legitimate nodes in the network.
\subsection{Denial of service (DoS)}
A denial-of-service (DoS) attack is any such situation when a legitimate node is denied access to network resources or systems. DoS attack can stop nodes from sending messages to reporting gateways, reject network joint requests and drop messages due to deliberate interference in the medium.\\

The other types of attack that are mentioned in the Table \ref{tab:ATtackTable} are \textit{selective forwarding}\cite{YuSelectFwd2006}, \textit{blackhole attack} \cite{tseng2011survey,kaur2017detection} \textit{sinkhole attack} \cite{krontiris2007intrusion}, \textit{Sybil attack} \cite{NewsomeSybil2004,PathanSybil2006 } and \textit{Hello flood} \cite{gill2018detection}. These attack types are exclusively applicable to multi-hop WSNs and adhoc networks as they manipulate routing information in multi-hop sensor network scenarios.
However, the existing LPWAN technologies (i.e., Sigfox and LoRa ) have star topology. The broad attack types that can affect long-range, low power single-hop networks are eavesdropping, jamming including selective jamming, collision, and wormhole with replay attack and ACK spoofing.

\section{Security threats unique to LoRa networks} \label{UniqueLoRa}
\label{uniqueLoRa}
This section elaborates on the unique security threats of LoRa networks. There are several survey papers~\cite{makhdoom2018anatomy, tomic2017survey, GuWWirelessSurvey2011, ChahidIoTProtocol2017} that have discussed the security threats in the wireless network. They have broadly divided the network into WLAN (using WiFi technologies)~\cite{afanasyev2010usage}, WSN (using Zigbee and Bluetooth)~\cite{suryadevara2012wireless, mackensen2012bluetooth} and Ad-hoc wireless networks~\cite{sarkar2016ad}. Normally these kinds of networks have a short-range (100 meters). WiFi has a high data rate (100 Mbps) while WSN Zigbee has low data rates (250 kbps). LoRa is meant to cover at least 1-2 miles and therefore has a very low data rate (50 kbps) but long on-air time. It makes LoRa susceptible to different DoS and MITM attacks.
\subsection{Selective jamming due to long on-the-air transmission time}
\begin{figure}[ht]
    \centering
\vspace{-0.3cm}
    \includegraphics[scale=0.5]{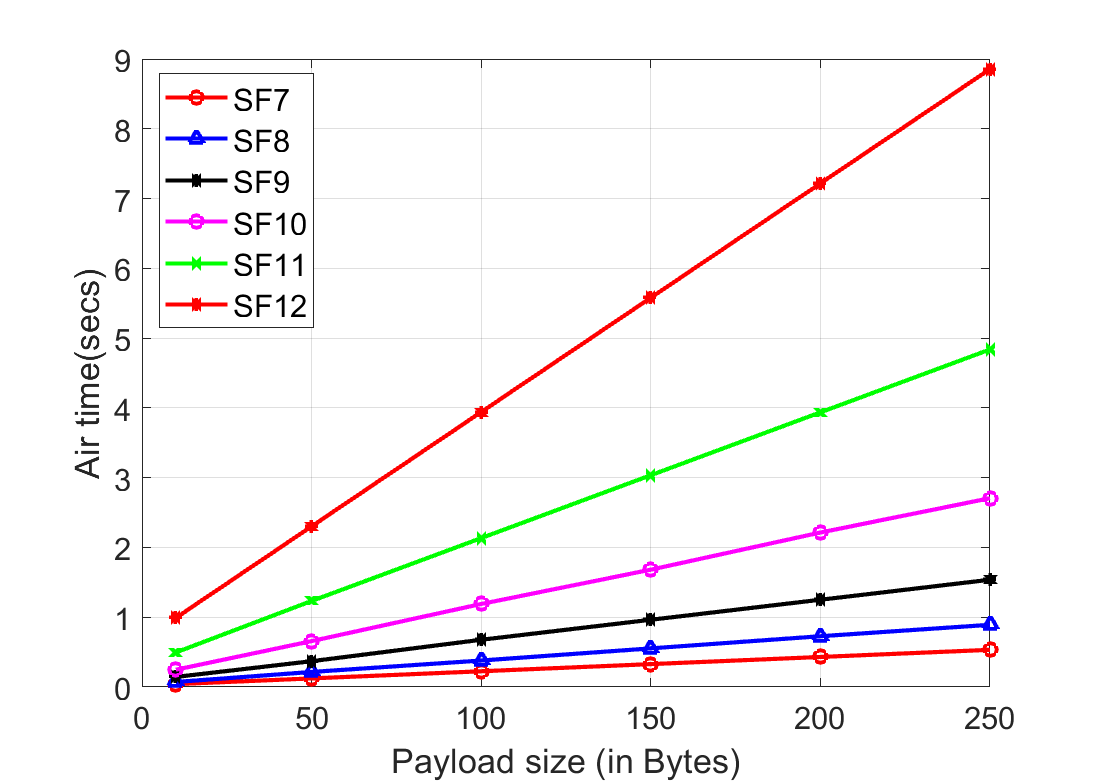}
    \caption{\textbf{Air time of packets as a function of SF and packet size}}
    \label{fig:LoRaOnAirTime}
    \vspace{-5 pt}
\end{figure}
Unlike other wireless technologies, like WiFi and ZigBee, LoRa employs an adaptive CSS (chirp spread spectrum) modulation scheme that can extend the communication range in a non-interference environment.  The data rate of LoRa networks is determined by SF and is updated by using an adaptive data rate (ADR) algorithm. Higher SF corresponds to lower data rates and in turn, long-range communication. Depending on the SF in use, the LoRaWAN data rate ranges from 0.3 kbps to 50 kbps (depending on the bandwidth), resulting in an on-the-air transmission time between 2.6 and 0.03 seconds, if the payload size is around 100 bytes. High transmission time creates opportunities for adversaries to perform an attack on the LoRa networks.

Fig. \ref{fig:LoRaOnAirTime} shows the air time of LoRa packets as a function of SF and payload size. The on-air time are calculated based on Eqns.\eqref{eq:OnAirtime}, \eqref{eq:Tpayload} and \eqref{eq:Tpreamble} provided in the LoRaWAN 1.1 Specification \cite{LoRaSpec}. 
\begin{equation}
\label{eq:OnAirtime}
T_{onAir} = T_{preamble} + T_{payload}
\end{equation}
where,
\begin{equation}
\begin{aligned}
\label{eq:Tpayload}
T_{payload} = 8 + max(ceil((8PL- 4SF+ 28 +16 -20H)/\\(4*(SF -2DE)))(CR +4),0)
\end{aligned}
\end{equation}
and 
\begin{equation}
\label{eq:Tpreamble}
T_{preamble} = (8 +4.25)Tsym 
\end{equation}
Here,
\begin{itemize}
    \item PL Is the number of payload bytes
    \item SF The spreading factor (between 7 and 12)
    \item H = 1 when no header is present, and H = 0 when no header is present.
    \item DE = 1 when the low data rate optimization is enabled, DE = 0 for disabled.
    \item CR is the coding rate from 1 to 4
    \item $T_{sym}$ is the symbol duration that depends on SF and bandwidth
\end{itemize}

Another unique characteristic of LoRa networks is that the difference between the lowest and highest data rates is vast, resulting in a significant difference between the on-the-air transmission time. The nodes that are close to the gateways can use a high data rate and a short on-the-air transmission time; on the contrary, the nodes that are far away from the gateways have to use a low data rate and a long on-the-air transmission time. As a consequence, the end devices in the same LoRa network may experience different levels of risk from the attacks of adversaries.

An attacker can make use of the long on-air time to launch a selective jamming attack \cite{aras2017exploring} when the attacker can read the physical message header (which is not encrypted) and jammed based on the jamming policy. Selective jamming not only reads through the preamble but also the message header. Thus attackers can listen on the channels, target a particular device or traffic class and then jam selected messages~\cite{proano2010selective}.
In order to selectively jam LoRaWAN messages, an attacker has to perform the following steps. It first detects a LoRaWAN packet. It aborts receiving if the received content triggers the jamming policy (usually first 5 bytes).
If no, it immediately jams the channel.  Therefore, the jamming window is smaller than the general triggered jamming.
However, selective jamming can prevent critical messages from reaching the gateway, especially for those sensor devices that only transmit when the sensor state change happens (event-driven sensors). Selective jamming is implemented on a real test-bed with cheap hardware \cite{aras2017selective}.

 \subsubsection{Selective jamming with Wormhole attack}
 Classical wormhole attack requires two malicious devices in a wireless sensor network \cite{YHuWormhole2003,MpriyaWormhole2012}. One device records regular network messages and tunnels them to the other device through a low-latency network.  Generally, this kind of attack is meant for mesh networks where false routing information is created to drain out the energy of the network. The classical wormhole attack is not suitable for LoRaWAN that uses the star topology. However, there is no time-related information in the LoRaWAN message header and only loose timing requirement due to LoRa'’s long transmission time.
Therefore, messages can be recorded, jammed, and then can be replayed later to appear as a valid message as long as a message with a higher sequence number is received at the gateway.
In this kind of selective jamming and wormhole attack, two devices are involved and act as a sniffer and a jammer separately. The sniffer receives the messages and decides whether to jam as per normal jamming policy. If the decision to jam is made, it signals to the jammer using a low latency network to jam the message immediately. The sniffer constantly sniffs the network. In order to carry out this kind of attack, the following steps are followed by the sniffer and the jammer.

The sniffer first detects a LoRaWAN packet
and starts receiving and recording the packet. By receiving a packet, it sends a signal through a low latency (better than LoRaWAN) to the jammer. By receiving the signal from the sniffer, the jammer turns to the active mode.

By replaying the recorded regular messages while jamming, the attacker not only intercepts the state change alert messages but also makes it look like nothing out of the ordinary has happened. The window of opportunity in this kind of attack is even lower than selective jamming as there are two devices involved. Devices that are using high SF and therefore, long on-air transmission time are more vulnerable to this kind of attack.

\begin{figure}[t]
\begin{center}
    \includegraphics[scale= 0.50]{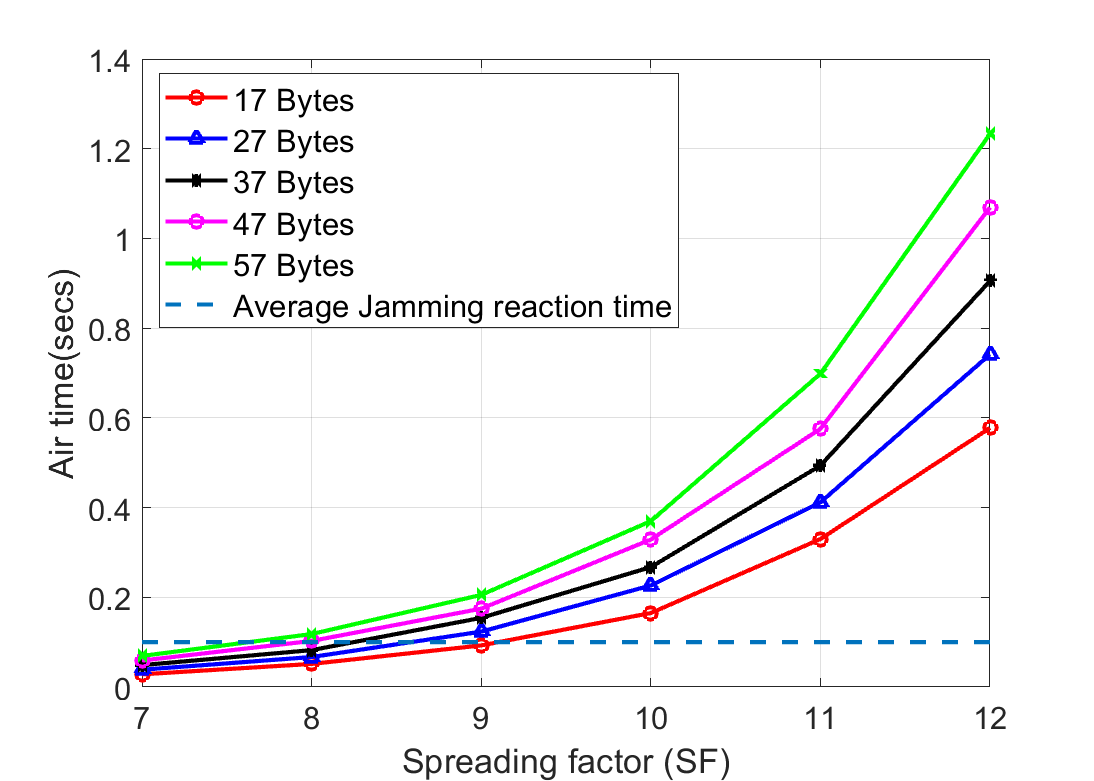}
\end{center}
       \caption{\textbf{Time available to jam for various packet sizes for each SF}}
    \label{fig:LoRaOnAirTime_2}
     \vspace{-15pt}
\end{figure}

Small scale experiments with wormhole attacks have been carried out in recent times as literature suggests \cite{aras2017selective}. The most critical aspect of a successful attack is that the reaction time of the jammer must be less than the on-air time of the LoRaWAN packet reduced by the message header bytes. In \cite{aras2017selective}, the mean reaction time of jammer is experimentally found to be 100 milliseconds. 
Fig. \ref{fig:LoRaOnAirTime_2} shows the time available to jam for various packet sizes for different SF when the mean reaction time is 100 milliseconds. It can be observed that LoRa devices that are on the edge of the network and using higher SF are more vulnerable to such kinds of jamming attacks.
This section shows that the threats to all LoRa devices are not the same as they are not equally vulnerable. End devices that are using low data rate or high SF have longer on-air time and prone to these kinds of jamming attacks. Jamming causes packet drop, the transmission of lost packets and overall higher energy consumption of sensor nodes. End devices adjust their data rate based on the distance from the gateway. There is no intelligence, neither in the gateway or in the end devices which might sit on top of the adaptive data rate that can adjust the transmission parameters to counter the attack.


The relatively long on-air-time transmission of the LoRa packet can result in \textit{ACK spoofing}. An attacker can prevent receipt of ACK packets through selective jamming of the end-device and would later replay previously recorded ACK to disrupt uplink messages.

\subsection{Collision attack due to ALOHA-type medium access control}
Although collision attacks and their mitigation techniques are not unique to LoRa networks, the challenge in LoRaWAN is to distinguish packet losses due to network congestion or deliberate injection of packets to create interference with legitimate packets. 

LoRaWAN standard specifies that LoRa devices use random time slots (pure ALOHA) to access the channel. Interference occurs when signals simultaneously collide in time, frequency, and SF. It can severely affect the network throughput. The restriction on the duty cycle of LoRa end devices also contributes to the overall network throughput performance. In the USA, each end device has to wait for 400 milliseconds after each transmission. Large SFs increases the time on not only air but also the off-period duration. The maximum data rate in such a network is limited by the on-air transmission time in each subchannel. In  \cite{adelantado2017understanding}, authors have shown that the throughput is limited by collision (pure ALOHA) when the traffic load and network size is low. With the increase in traffic load (total transmission rate) and network size, the throughout stabilizes because the duty cycle limitation restricts the LoRa end devices to increase the packet transmission rate. Different traffic loads (0-350 packets per hour per node) and the number of end devices of 250, 500, 1000, and 5000 were used. The payload size was 10 bytes, and the number of channels used is 3. The results demonstrate that the performance behavior of LoRa networks is unique as compared to WiFi and Zigbee.

Existing work in  \cite{georgiou2017low,AhoellerLoRaIssues2018} suggests that dense network deployment can adversely affect the network throughput. The authors in these papers have also suggested that in an overlapping LoRaWAN network scenario, the effect of the collision on network performance needs further investigation.
 
 On top of packet losses due to the collision, jamming attacks can aggravate the situation further by injecting malicious packets over a wide frequency band. Since LoRaWAN does not have any mechanism to sense channel before transmitting, it will be hard for LoRa networks to mitigate a jamming attack, especially in dense deployment scenarios.
 
 In \cite{PhamLoRaImage2018, PhamLoRaCSMA2018}, the author has proposed to use LoRa's channel activity detection (CAD) process to sense channel and reduce collisions. According to the application note from Samtech, the proprietary designer of LoRa transceivers, the CAD model is designed to detect a LoRa preamble on the radio channel with the best possible power efficiency \cite{SamtechCAD}. The LoRa radio receiver captures the LoRa preamble symbol of data from the channel. The LoRa radio current consumption during that phase is approximately 10 mA. The CAD duration is a function of the spreading factor (SF). The energy consumption to implement the CAD process for channel sensing is energy prohibitive and suitable for only Class B and C end-devices. Also, the author has not considered the scenario when there are packet losses. 
The author pointed out that the performance of channel sensing (CS) using CAD started to decrease beyond 1 kilometer and fails to detect channel activity many times during an ongoing transmission. 

In \cite{LayaByeAloha2016}, the authors have proposed the distributed queueing (DQ) algorithm for channel access that will mainly cater to the IoT network traffic and present itself as an alternative to ALOHA. The motivation for the authors stemmed from the wide variety of applications that use IoT networks for communication. Examples include structural health monitoring, asset tracking, automatic meter reading, and power grid protection and control. In these application scenarios, the end devices usually remain in sleep mode to save energy and only wake up to transmit bursts of data.


 
\begin{figure}[t]
    \centering
\vspace{-0.3cm}
    \includegraphics[scale=0.4]{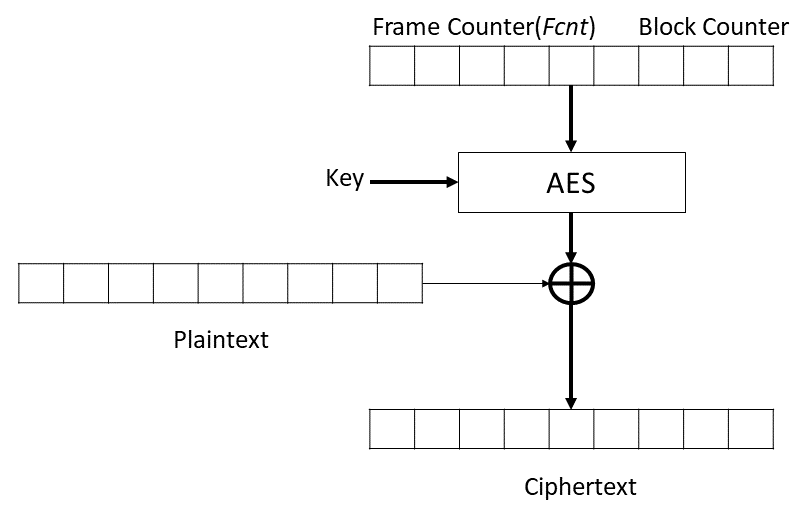}
    \caption{\textbf{LoRaWAN uses AES to encrypt messages }}
    \label{fig:AESCTR}
     \vspace{-10 pt}
\end{figure}

\subsection{Encryption Key management issues in LoRaWAN}
LoRaWAN employs NwkKey (AES-128) and AppKey (AES-128) to protect data from a man-in-the-middle attack. LoRaWAN implements AES in counter mode. The operation is shown by a block diagram in Fig.\ref{fig:AESCTR}.
Instead of a random number, LoRaWAN uses the frame counter as nonce along with a monotonically increasing block counter to create the cipher stream. This cipher stream goes as input to the AES encryption algorithm block. The key and the cipher stream are used to encrypt the plain text to create the ciphertext. If the counter block values are repeated with the same key, then the same keystream is used to encrypt consecutive packets, and confidentiality guarantees are void \cite{AESCTR2004}. In LoRaWAN, when the packet or frame counter is reset, while the key is not changed, the block cipher recreates the same \textit{key} values. Some previous studies~\cite{yang2017lorawan,10.1007/11894063_8,10.1007/978-3-540-45203-4_23,10.1007/978-3-540-85893-5_4} have shown that it is possible to compromise this encryption method by capturing consecutive packets from a device. If an eavesdropper captures two on-air ciphertext streams using the same key-stream, it can get the XOR-ed of two plaintexts by canceling out the key. It is then straightforward to separate them \cite{AESCTR2004}. 


We infer from this section that an attacker can affect the integrity and availability of the LoRa network even without deciphering the on-air packets.  Using high power jammer, it can cause interference of the channel, resulting in packet losses. It can also endanger the integrity of packets by affecting the randomness of specific numbers that are supposed to guarantee message integrity.


\section{Current solutions in the context of LoRaWAN }
\label{LoRaSolutions}
Several techniques have been proposed to detect and counter different attack scenarios, especially in the context of jamming \cite{MpitAJammingSurvey2009}. It is argued in that paper that anti-jamming measures are not usually considered in wireless sensor network design, but a formidable attack can seriously jeopardize high layer security mechanisms. While there are pro-active measures to mitigate signal interference due to jamming, detection of an attack should be the first step as attack models may not be known in advance.

In this section, we discuss the state-of-the-art security countermeasures in wireless sensor networks and their suitability for LoRa networks. 

\subsection{ Detection and mitigation against jamming attack }
Detecting jamming attacks is crucial because it is the first step towards building a secure and dependable IoT network. As mentioned in section \ref{UniqueLoRa}, the LoRa network can suffer from packet losses due to collision due to the random medium access policy. A packet can also be lost due to poor radio link quality. An attacker can also inject short packets in the network to increase the collision rate. Therefore, it is hard to detect packet losses due to the nature of the wireless channel access mechanism from an actual jamming attack. In \cite{XuJammig2005}, authors have proposed the idea of consistency check to detect an attack from regular network congestion. They propose two enhanced algorithms using signal strength for consistency check and location information for consistency check separately. In the first method, the packet delivery ratio (PDR) is combined with the measured signal strength to build four different scenarios. Results show that this method has improved jamming detection accuracy. In the location information method, a node finds its jamming status by checking its PDR with the one that is consistent with its location. However, if the node is mobile, it has to advertise the location to get the consistency checked periodically. 

 Another well-known detection technique is proposed in \cite{Zhoujamming2005} but is primarily designed for TDMA protocols. It involves the exchange of packets between a transmitter and a receiving node to detect the amount of interference and subsequent propagation of the interference information to the neighboring nodes. This information is used to figure out the collision scenarios in the network. 
 
The JAM or jamming area service for sensor networks is a \textit{reactive countermeasure} to detect and map the jamming region and then re-route incoming traffic through the unaffected region \cite{WoodADJAMarea2003}. 

\subsubsection{Proposed solution and comments}
A high power wideband continuous jamming attack can disrupt the service over multiple frequency bands but are easy to detect. However, a selective jammer that only disrupts communications for specific packets is difficult to detect and deter. A combination of proactive and reactive measures is required to counter this kind of jamming attack.

The \textit{location-aware} consistency check method can be used by LoRaWAN to detect a jamming attack. However, this method needs the support of the location service of the wireless network. The traditional way to locate a device is to install a GPS sensor within the device. LoRaWAN solutions by design are low-power, and long battery life (up to 10 years).  However, GPS sensors are power-hungry and can reduce the battery life of a LoRaWAN sensor by a factor of 40-50. However, with LoRaWAN geolocation service \cite{LoRaGeolocation2018}, it is possible to triangulate the position of the node using timestamps when it is connected to 3 or more LoRa gateways. LoRaWAN can combine the signal strength and the \textit{location information consistency check} method to detect a jamming attack.

The detection method that is proposed in \cite{Zhoujamming2005} can work for adjacent nodes and will be vulnerable to any external jamming attack. Further, this technique also uses network bandwidth and energy that is in the premium. LoRaWAN protocol is not well suited to handle high overheads, and LoRa end devices only wake when there is information to transmit.

The JAM may work for a multi-hop scenario while LoRa network only supports single-hop topology.

\subsection{Game-Theoretic approach against jamming attack}
The relationship between legitimate transmitters and jammers is antagonistic, and they are capable of controlling their actions intelligently. Therefore, game theory \cite{GameTheory} is a natural tool to model and analyze their behaviors in wireless
networks.
In \cite{SdoroGameTh2015,MLabibGameTh2015,NNamvarGameTh2016}, the authors have proposed to countered the jamming attack with game-theoretic approach. The \textit{Colonel Blotto Game} model is used to allocate resources (power) among the nodes that are affected by the jamming attack so that the network performance in terms of successful packet transmissions and threshold bit error rate (BER) is maintained. The nodes periodically report the interference level to the gateway. The nodes in the mesh network are classified based on their number of connectivity. Nodes with a high number of connectivities are assigned more bits to report to the gateways, therefore allocating more control bandwidth. In this jamming mitigation technique, it is assumed that each IoT node has a interference sensor that relays the measured interference level to the gateway through a common control channel. The timing channel is also used to maximize information throughput (in bits/sec) in the presence of a jammer\cite{DoroSGameTHeorySilence2016}. 

In \cite{GaoYBimatrixGameTheory2018}, a bimatrix game framework is developed for modeling the interaction process between the transmitter and the jammer, and the sufficient and necessary conditions for Nash equilibrium (NE) strategy of the game are obtained under the linear constraints.

In \cite{LiYStacklberg2014}, the authors have modeled the interaction between transmitters and a jammer as a \textit{Power Control Stackelberg game} in co-operative anti-jamming communications. In a co-operative wireless network, a relay node helps the source send a message in the presence of a smart jammer. The relay node improved the signal to interference plus noise ratio (SINR) and led to a higher throughput. The legitimate transmitter and the jammer can control transmission power independent of each other. 

It is pointed out in \cite{ZhangMultiStackelberg2018} that in the presence of high power jammer, power allocation may not be enough to counter jamming. The authors have considered a multichannel and multi-user scenario. The utility of the users is defined as the tradeoff between the system total throughput reward and single user’s power consumption. Therefore, the jammer engages with one user and worsen the communication, but the overall throughput can improve as the jammer is busy with one user. 

\subsubsection{Proposed solution and comments}
The control channels (time and bandwidth) are used to maximize the information exchange between nodes and the Gateways. In the context of LoRaWAN, The packet transmission time $T_{P}$ that is assumed in the solutions is in the order of $\mu$seconds. However, this method is not effective in LoRa as the $T_{P}$ can go as low as 40 milliseconds. The utility function under the constraint of energy never reaches 1.

In co-operative anti-jamming communication using \textit{Stackelberg game} model, it is assumed that there is limitless power in both the transmitter and the jammer. In a LoRa Network, the SF can only be controlled, and there is finite power in the end-devices. However, if LoRa end-nodes can control power beside SF, the co-operative Stackelberg game model can be used for interaction between jammers and the LoRa network.

In the context of LoRa network, power control may not be feasible, but \textit{dummy} users can be created to distract the jammer while the regular communication between the \textit{real} nodes and the gateway continues.

\subsection{Reinforcement learning-based anti-jamming techniques}
Traditionally, spread spectrum techniques have been used as anti-jamming methods in wireless communication. However universal reconfigurable radio peripherals can be used by jammers to block most frequency channels and interrupt legitimate communication. In \cite{RestFML2018}, the authors have argued that as IoT networks have been pervading every aspect of our lives, security attacks are myriad and have become even more challenging than ever before. Therefore, machine learning (ML) and software-defined radios and networks can provide reconfigurability and intelligence to the IoT devices and gateways.

From the reconfigurability aspect, Cognitive radio (CR) technology enabled by Software-defined Radio peripherals provides an excellent platform to implement learning algorithms to counter jamming. The learning algorithms can help the CR to find the frequency space that has minimal signal-to-interference ratio and improves the utility of the secondary user (SU).

For a successful implementation of supervised machine learning algorithms, it is essential to have a clear and consistent understanding of inputs (i.e., data), the states of the attack and the outputs. The state of the attack can be binary, i.e., the good IoT network state or the bad IoT network state. To build accurate classification models, the ML algorithm is required to be trained with a considerable number of examples. For all practical reasons, it may not be a feasible option to get training data sets.

Reinforcement learning fills the gap between supervised learning where the algorithm is trained on the correct answers given on a data-set, and unsupervised learning that exploits the similarity in the data-set to cluster it\cite{MarslandBook2}. The main advantage of using reinforcement learning to detect and counter malicious attacks is that no advance knowledge of the jamming attack model is required. It does not rely on training data sets but learns to find the right moves in order to optimize rewards according to the current state. 
\subsubsection{Proposed solution and comments}
The Deep Reinforcement Learning (DRL) framework can be used to learn a model-free neural network, which takes the LoRaWAN network context as input and generates the best countermeasure to defend against underlying attacks. DRL is a promising machine learning approach,
which instructs an agent to accomplish a task by trial and error in the process of interacting with the LoRaWAN network. The DRL inference results can tune the setting of corresponding end devices, like transmission power, on-air transmission time (determined by spreading factor) and error-correcting code. Trust-able cryptography schemes encrypts the results and transmitted to the end devices.

Four key elements of our DRL framework will be defined to describe the learning process of DRL, i.e.,
\textit{state}, \textit{action}, \textit{policy} and \textit{reward}. The state \textit{s} defines the input of an agent, referring to the representation of the LoRa network state, which can be quantified by passively received transmission data from every end device, like packet reception ratio and signal to noise ratio. The policy $\pi$ takes the LoRa network state as input to generate an action. It learns a mapping from every possible state to action according
to the past experience. In our framework, the policy is implemented as a deep neural network (DNN). The
action \textit{a} affects the LoRa network. Every action gets feedback from the LoRa network. According
to the feedback, we calculate a \textit{reward r(s, a)}, which indicates how good or bad does an action change the
LoRa network given a specific state s. Based on the reward, a value function \textit{Q(s, a)} is defined to update
the policy of the agent. The Q value reflects the long-term effect of an action, e.g., if an action has a high Q
value, the parameters of the DNN agent will be updated to favor that action.

The DRL agent learns to defend against different DoS attacks by training with a specific policy, supposing we can produce
enough transition samples ($s_{t}$, $a_{t}$, $r_{t}$, $s_{t+1}$) on a testbed. The agent first perceives a state $s$ and generates an action $a$ by running the policy $\pi$. Then, the agent obtains a reward $r$ calculated by the reaction of the LoRaWAN and updates the policy based on the estimate of $Q(s, a)$. In this way, the agent and the LoRa network interact with each other to modify the policy. After sufficient iterations, the agent learns a stable policy. Besides, after each online inference, the agent can also use the above training process to
update the policy of the DNN agent incrementally based on the new traffic data.

To incorporate DRL into practical DoS attack countermeasures, there are a set of challenges, including context-aware DRL input representation, light-weight agent, efficient agent training process and scalable DRL framework for new attacks and dynamic network topology. 

- First, to enable a more accurate and efficient defense, the network context must have a fine-grained representation. The statistical information, like packet reception rate, RSSI, and the location of the sender, will be represented in a vector and will be taken as an input to the agent.

- Second, the output of the agent is the transmission setting of all end devices. For large-scale LoRaWAN networks, we may have more than thousands of end devices. The output size is large, and the action space
is huge, resulting in long inference time. One successful implementation of DRL is Deep Q-Network (DQN) \cite{hester2018deep}. DQN algorithm combines Q-learning with a DNN \cite{van2016deep}.

Q-Learning is a field of reinforcement learning paradigm where the expected reward is calculated from the current state and each possible action that can be taken separately in the action space. Q-Learning can learn the optimal policy by trials in the Markov Decision Process (MDP). It is pointed out in \cite{HanGRML2017} that the Q-learning algorithm can suffer from slow convergence to the optimal policy if the state space and action sets are large, resulting in degraded anti-jamming performance. They have proposed DQN that was developed by Google DeepMind \cite{Deepmind2016}. It exploited the deep convolution neural network (CNN) to reduce the dimensions of the Q-learner and accelerate the learning process. In \cite{SlimeniCRMDP2015}, authors have also used MDP and reinforcement learning to mitigate jamming in Cognitive radio. They replaced the OFF-policy characterizing the standard Q-learning algorithm with an ON-policy so that the cognitive radio takes the best action corresponding to the maximum of the quality value, instead of trying random action. This method will take care of any new jammer as and when they emerge in the network.

LoRa network can leverage the use of different SF to create the state ($s$) and action ($a$) space where it has SF from 7 to 12 to choose from \cite{AugustinLoRa2016}. The SF factor controls the data throughput and therefore, the on-air time of a packet for a given frequency band (125 kHz, 250 kHz, and 500 kHz) and packet size. The LoRa end devices will probably be non-configurable due to constraints on stored power and computational capabilities. However, the LoRa gateway can be used as a reconfigurable platform that is enabled with reinforcement learning capabilities. It can be used to make decisions for the LoRa end-devices. 

\subsection{Power and rate adaptation approach}
In \cite{PelechrinisJamming2009}, authors have proposed ARES (Anti-jamming Reinforcement System) for  IEEE 802.11 compliant networks to counter jamming attacks. They combine the rate adaptation and power control to ensure communication in the presence of radio interference caused by a jammer. The rate control module chooses the data rate depending on the channel conditions and the jammer characteristics.  The power control module adjusts the clear channel assessment (CCA)  \cite{fonseca2007clear} threshold of the IEEE 802.11 standard to facilitate the transmission and the reception of valid packets during jamming.
\subsubsection{Proposed solution and comments}
Although LoRa has no CCA mechanism to sense channel state, it can beacon packets during network setup to determine the PDR without any jamming interference and periodically compare the current PDR to control the data rate (with SF) and power (if possible).
\subsection{Spread Spectrum based methods against jamming attacks}
 As a \textit{pro-active countermeasure} in the link-layer MAC protocols, it is suggested to shorten the packet size, increase the duty cycle, use time slots for packet transmission (TDMA), or use different spread spectrum techniques (FHSS, DSSS or UWB). The spread spectrum techniques use pre-shared keys (either the hopping sequences or the spreading codes) between the communication partners. Such a technique enables nodes to spread signals in time and frequency or both \cite{CHangFreHop2014}. Several variants of frequency hopping (FH) have been proposed in the literature to counter malicious attacks. Un-coordinated frequency hopping does not provide a shared key but is dependent on the maximum time for two nodes to rendezvous on a frequency and the randomness (entropy) of the frequency sequence. In \cite{GopalFreHop2017}, authors have used chaotic frequency hopping for anti-jamming communication in a body area network. Authors in \cite{HannonFreqHop2016} has proposed frequency hopping based on jamming statistics. In paper \cite{LeeQUROM2012}, the authors have introduced a random spread-spectrum based wireless communication scheme that can achieve both fast and robust data transmission. In the proposed scheme, Frequency Quorum Rendezvous (FQR), it is guaranteed that two random hopping sequences will rendezvous within a limited time.  

The general direct sequence spread spectrum (DSSS) uses pre-known spreading codes based on a shared key to spreading the message. However, an attacker can get hold of the key and spread any random message and send it to the receiver, thereby confusing it completely. To overcome this situation, the authors in \cite{AlagilDSSS2016} have proposed the use of randomized positioning DSSS to avoid a jamming attack. Authors in \cite{WoodADDEEJAM2007} have proposed DEEJAM, a MAC layer protocol to defend against jamming using IEEE 802.15.4 compliant hardware. It uses frame masking, frequency hopping, and packet fragmentation to counter a jamming attack. 
\subsubsection{Proposed solution and comments}
The performance of randomized approaches is measured by the latency or delay in the connection setup phase that depends on the probability of encounters. Currently, no LoRa node uses this scheme to communicate. Under intense jamming conditions, it may not be a feasible solution as nodes may have to wait forever to communicate. 

\subsection{Encryption key management against Replay attack}
The authors in \cite{KimReplayAttack2017} have elaborated on the limitation of using DevNonce to counter replay attacks that are discussed in Section \ref{Intro}. On the other hand, the limitation of the LoRaWAN system to use the same AppKey for a lifetime can be a serious problem as \textit{key leakage} can cause node capture and side-channel attacks \cite{kimdualkey2017}. Under the current LoRaWAN key usage scheme, the network server generates both the session keys (NwkSKey and AppSKey). If the network server is compromised, it can use the AppSKey to intercept the application layer data. In \cite{kimdualkey2017} the authors have proposed the use of Dual Key-Based Over-the-Air Activation where they have introduced NwkKey as a new network server access key. It is separated from the original AppKey which will only be used to access the application server.
\subsubsection{Proposed solution and comment}
In this proposed join procedure, NwkSKey and AppSKey are generated from the NwkKey and the AppKey, respectively. This kind of key management approaches can create two distinct fire-wall where deciphering one key will be not enough to launch an attack.

\subsection{ Detection of wormhole attack using packet leashes}
In \cite{hu2003packet}, the authors have proposed a technique called \textit{packet leashes} to defend against wormhole attacks in wireless networks. In this detection method, there is an upper bound on the distance or time of travel of the packet. If that maximum distance or time is exceeded, then a wormhole attack alarm is raised.
In \cite{LazosLWormhole2005}, the proposed solution involves a small fraction of network nodes to have location information (using GPS) to precisely know the distance between the nodes to restrict the packet’s maximum transmission distance. In \cite{Hu2004UsingDA}, directional antennas can also be used to mitigate the wormhole attack.  
\subsubsection{Proposed solution and comments}
The approaches that are discussed in \cite{hu2003packet, Hu2004UsingDA, LazosLWormhole2005} require accurate time synchronization or timing measurement, or to transmit maximum power in a particular direction.  In the context of resource constraint LoRa nodes, the applicability of the solutions is ineffective. These methods are also not resource-efficient as they involve using control bandwidth to communicate to detect an attack. The nodes are also required to listen for interference from malicious sources. However, selective jamming with Wormhole attack only affects individual packets by using \textit{sniff and jam} method. The proposed methods will be rendered ineffective in these scenarios.

\subsection{Information theoretic approach against eavesdropper}
In a traditional secure communication setup, a transmitter wants to send a message securely to a legitimate receiver without an eavesdropper reading the message. From the information-theoretic model, such a communication channel is modeled as a wire-tap channel model where we have a broadcasting transmitter, a legitimate receiver, and an illegal receiver \cite{OggierFSecreacyChannel2008}. Perfect secrecy is achieved if a transmitter can reliably and confidentially send a message to the intended receiver without exposing any bit of information to an eavesdropper. The secrecy capacity has been studied in SIMO, MISO and MIMO setups in fading channels in the recent past.

In \cite{BarrosJSecrecy2006}, the authors have defined the \textit{secrecy capacity} in terms of \textit{outage probability} and characterized the maximum rate at which the sniffer or eavesdropper will not be able to decode the message. In \cite{KhistiAJAmmingMIMO2007, CumananKSecreacy2014}, the authors have considered multi-antenna scenarios in the transmitter, receiver, and the sniffer. 

The basic premise of the research outcome is that the noise signal can be injected by the transmitter to degrade the received signal quality of the potential eavesdroppers \cite{CumananKPhysicalJaming2017}. This method can also deter potential jamming attackers as it will be difficult to sniff the transmitting channel to decode the preambles before itself launching a jamming attack. 
The authors have provided a solution for power minimization and secrecy rate maximization, for a MIMO secrecy channel in the presence of a multiple-antenna eavesdropper.

\begin{table*}[ht]
  \centering
  \caption{Summary of possible solutions to attacks in LoRaWAN}
  \label{tab:AttackSol}
  \begin{tabular}{|p{7em}| c|p{5em} |p{13em}|p{14em}|}
  \hline
    Attack type & Affected layer & Counter-measure  & Techniques & Remark\\
    \hline
    \multirow{3}{*}{Jamming attack} & \multirow{3}{*}{Physical layer} & Detection & Consistency check with PDR, signal strength and location information &  LoRaWAN provides location service using three or more LoRa gateways \\ 
    \cline{3-5}&
     & Re-active measure &  Game-Theoretic approach using \textit{dummy} nodes as a player while the real nodes transmit packets &  LoRaWAN can deploy extra nodes to distract an attacker \\ 
     \cline{3-5}&
     & Re-active measure &  Reinforcement learning to determine the best policy to counter an attack and minimize jamming affect & Use different SF from the SF pool to control the data rate and link quality \\ 
     \hline
    
    \multirow{2}{*}{Collision attack } & \multirow{2}{*}{MAC layer} & Pro-active measure &  Different frequency hopping (FH) and spread spectrum (SS) techniques to increase randomness and uncertainty for an attacker & LoRaWAN can adopt hybrid FH and SS methods to counter selective jamming attacks    \\ 
    \hline
    
    \multirow{2}{*}{Replay attack} & \multirow{2}{*}{Multi-layer} & Pro-active measure & Dual-key based authentication during network joining &  Separate server access keys for the network and the application servers generated separately from Network and Application keys provided by the vendor     \\ 
    \hline
        \end{tabular}
\end{table*}
In \textit{transmit jamming}, the assumption is that the legitimate transmitter-receiver pair is aware of their channel state information (CSI) while the CSI of the attacker is not known in advance as it is not realistic. The transmitter splits the transmitting signal into two sections. Based on the CSI of the receiver, the transmitter will beam the information-bearing signal towards the receiver. The other section is the noise signal that is orthogonal to the intended received signal. The purpose is to degrade the quality of the signal received by the sniffing attacker.

In \textit{receiver jamming} method, the receiver confuses the eavesdropper by transmitting the jamming signal. In this method, the legitimate transmitter re-transmits the message certain times, and the receiver randomly jams the message. Such a scheme will make the sniffer unable to decode the message. 
\subsubsection{Proposed solution and comments}
 The \textit{transmit jamming} technique requires careful distribution of transmit power between the receiver and sniffer as well as overall significant energy consumption. LoRa does not support multi-antenna propagation, and the use of transmitting jamming will end up decreasing the SNR at the intended receiver.

The \textit{receiver jamming} operation is bandwidth and energy inefficient and will not suite LoRa devices that are constraint by energy and bandwidth. Even LoRaWAN access protocol, i.e., pure ALOHA, will not be able to support this kind of jamming attack mitigation technique and end up losing too many packets due to collisions.

\subsection{Detecting passive eavesdropper with leakage RF signal}
 In paper \cite{ChamanEavesdropper2018}, the authors designed and built a device called Ghostbuster that can detect RF leakage of passive receivers that are buried within the current transmission. Results show that their device can detect eavesdroppers with more than 95\% accuracy up to 20 feet away.
 \subsubsection{Proposed solution and comments}
Such a solution against eavesdropping is suitable for short-range communication. But a LoRa network is deployed for long-range communication, and an eavesdropper can be located anywhere within a radius of 1-2 miles. However, the LoRa network can set up dummy gateways to detect eavesdroppers.

\subsection{Summary of possible solutions}

Based on the discussions of the previous sections, we have summarized the possible solutions in Table \ref{tab:AttackSol}.
It will require a cross-layer approach to detect and take action against the different attack vectors.



\section{Conclusion and Future Research Direction}
The LoRa network has gained prominence due to its long-range and low power operation in the IoT application domain. It has been used for data collection and processing from very general to very critical application scenarios. Intentional radio interference to disrupt legitimate communication is detrimental to the full-blown use of the LoRa network. There exist techniques to address the different DoS attacks. A DoS attack can cause serious message integrity and confidentiality issues. A DoS attack can jam specific packets, and even eavesdrop to launch a replay attack. Conventional countermeasures are not sufficient to deal with these attacks. Specific approaches using game theory can be applied to the LoRa network to thwart jammers that disrupt on-going communication. Since an attack model may not always be available, online reinforcement learning algorithms can be exploited to tackle jamming. Efficient learning is a trade-off between exploration and exploitation. It is required to develop a deep reinforcement learning framework to defend against DoS attacks. The proposed solutions
should meet the constraints of LoRa end devices and should be evaluated by a comprehensive set of performance metrics, including effectiveness, memory cost, processing time, and power consumption. Overall, there are significant research scopes to design practical LoRa networks by integrating security solutions into networking system development.

\bibliographystyle{IEEEtran}

\bibliography{bibliography}

\end{document}